\documentclass[12pt]{article}
\usepackage{amsmath,amssymb,amsthm}

\newtheorem{theorem}{Theorem}[]
\newtheorem{lemma}[theorem]{Lemma}

\begin{document}

\title{Weighted Efficient Domination for $P_5$-Free Graphs in Linear Time}

\author{
Andreas Brandst\"adt\footnote{Institut f\"ur Informatik,
Universit\"at Rostock, Germany. e-mail:
{\texttt ab@informatik.uni-rostock.de}}
}

\maketitle

\begin{abstract}
In a finite undirected graph $G=(V,E)$, a vertex $v \in V$ {\em dominates} itself and its neighbors.
A vertex set $D \subseteq V$ in $G$ is an {\em efficient dominating set} ({\em e.d.} for short) of $G$ if every vertex of $G$ is dominated by exactly one vertex of $D$. The {\em Efficient Domination} (ED) problem, which asks for the existence of an e.d. in $G$, is known to be NP-complete for 
$P_7$-free graphs but solvable in polynomial time for $P_5$-free graphs. 
Very recently, it has been shown by Lokshtanov et al. and independently by Mosca that ED is solvable in polynomial time for $P_6$-free graphs. 

In this note, we show that, based on modular decomposition, ED is solvable in linear time for $P_5$-free graphs.  
\end{abstract}


\section{Introduction}

Let $G=(V,E)$ be a finite undirected graph with $|V|=n$ and $|E|=m$. A vertex $v \in V$ {\em dominates} itself and its neighbors. A vertex subset $D \subseteq V$ is an {\em efficient dominating set} ({\em e.d.} for short) of $G$ if every vertex of $G$ is dominated by exactly one vertex in $D$.
Note that not every graph has an e.d.; the {\sc Efficient Dominating Set} (ED) problem asks for the existence of an e.d.\ in a given graph $G$.
If a vertex weight function $\omega: V \to \mathbb{N}$ is given, the {\sc Weighted Efficient Dominating Set} (WED) problem asks for a minimum weight e.d. in $G$, if there is one, or for determining that $G$ has no e.d. The importance of the ED problem for graphs mostly results from the fact that ED for a graph $G$ is a special case of the {\sc Exact Cover} problem for hypergraphs (problem [SP2] of \cite{GarJoh1979}); ED is the Exact Cover problem for the closed neighborhood hypergraph of $G$.

For a graph $F$, a graph $G$ is called {\em $F$-free} if $G$ contains no induced subgraph isomorphic to $F$.
Let $P_k$ denote a chordless path with $k$ vertices. $F+F'$ denotes the disjoint union of graphs $F$ and $F'$; for example, $2P_2$ denotes $P_2+P_2$.

Many papers have studied the complexity of ED on special graph classes - see e.g. \cite{BraEscFri2015,BraGia2014,BraMilNev2013,Milan2012} for references. In particular, a standard reduction from the Exact Cover problem shows that ED remains NP-complete for $2P_3$-free (and thus, for $P_7$-free) chordal graphs. In \cite{BraEscFri2015}, it is shown that for $P_6$-free chordal graphs, WED is solvable in polynomial time. Very recently, it has been shown by Lokshtanov et al. \cite{LokPilvan2015} that ED is solvable in polynomial time for $P_6$-free graphs in general; independently, in a direct approach, also Raffaele Mosca \cite{Mosca2015} found a polynomial time solution for WED on $P_6$-free graphs. 

\medskip

A set $H$ of at least two vertices of a graph $G$ is called \emph{homogeneous} if $H \not= V(G)$ and every vertex outside $H$ is either adjacent to all vertices in $H$, or to no vertex in $H$. Obviously, $H$ is homogeneous in $G$ if and only if $H$ is homogeneous in the complement graph $\overline{G}$.
A graph is {\em prime} if it contains no homogeneous set. A homogeneous set $H$ is \emph{maximal} if no other homogeneous set properly contains $H$.
It is well known that in a connected graph $G$ with connected complement $\overline{G}$, the maximal homogeneous sets are pairwise disjoint and can be determined in linear time using the so called {\it modular decomposition} (see e.g. \cite{McCSpi1999}). 

\begin{theorem}[\cite{BraGia2014,BraGiaMilNev2014}]\label{thm:prime}
Let ${\cal G}$ be a class of graphs and ${\cal G}^*$ the class of all prime induced subgraphs of the graphs in ${\cal G}$. If the $($W$)$ED problem can be solved for
graphs in ${\cal G}^*$ with $n$ vertices and $m$ edges in time $O(T(n,m))$, then the same problem can be solved for graphs in ${\cal G}$ in time $O(T(n,m)+m)$.
\end{theorem}

For $v \in V$ let $deg(v)$ denote the degree of $v$ in $G$. Let $\delta(G)$ denote the minimum degree of a vertex in $G$. 
The modular decomposition approach leads to a linear time algorithm for WED on $2P_2$-free graphs (see \cite{BraMilNev2013}) and to a very simple $O(\delta(G) m)$ time algorithm for WED on $P_5$-free graphs (a simplified variant of the corresponding result in \cite{BraMilNev2013}). 

In \cite{BraGiaMilNev2014,BraMilNev2013}, WED is solved in time $O(\delta(G) m)$ for $P_5$-free graphs but it remained an open problem whether it can be solved in linear time. In this note we show that, based on modular decomposition, WED is solvable in linear time on $P_5$-free graphs.

We show that the WED problem can be solved in linear time for $P_5$-free graphs. This is based on some properties of $P_5$-free graphs with e.d.; by Theorem \ref{thm:prime}, we can restrict ourselves to prime graphs. 

\section{Structural properties}

A {\em thin spider} is a split graph $H=(V,E)$ with partition $V=C \cup I$ into a clique $C$ and an independent set $I$ such that every vertex of $C$ has exactly one neighbor in $I$ and vice versa. We know already:  

\begin{theorem}[\cite{BraMilNev2013}]\label{ED2P2fr} 
If $G$ is a prime $2P_2$-free graph then $G$ has an e.d. if and only if $G$ is a thin spider. 
\end{theorem} 


Let $G=(V,E)$ be a prime $P_5$-free graph with e.d. $D=\{d_1,d_2,\ldots,d_k\}$, $k \ge 2$ (note that $k=1$ is impossible since a prime graph does not have a universal vertex). Assume that $G$ is not $2P_2$-free. Let $R:=V \setminus D$ and let $R_i:=N(d_i) \cap R$. Since $D$ is an e.d. of $G$, $R$ is partitioned into $R_1,\ldots,R_k$. 
We claim:
\begin{equation}\label{Ndiclique}
\mbox{For every } i \in \{1,\ldots,k\}, R_i \mbox{ is a clique.}
\end{equation}

{\em Proof.} Suppose that $R_1$ is not a clique; let $a,b \in R_1$ with $ab \notin E$. Let $Q_{ab}$ denote the co-connected component in $G[R_1]$ containing $a,b$. Since $Q_{ab}$ cannot be a homogeneous set in $G$, there is a vertex $c \notin Q_{ab}$ distinguishing a non-edge in $Q_{ab}$, say $ca' \in E$ and $cb' \notin E$ for $a',b' \in Q_{ab}$ with $a'b' \notin E$ (and thus $c \notin R_1$; say $c \in R_2$) but then $b',d_1,a',c,d_2$ induce a $P_5$  which is a contradiction showing 
(\ref{Ndiclique}).  
\hfill $\diamond$

\medskip

We claim:
\begin{equation}\label{edgeRiRj}
\mbox{For every } i,j \in \{1,\ldots,k\}, i \neq j,  \mbox{ there is an edge between } R_i \mbox{ and } R_j.
\end{equation}

{\em Proof.} Suppose that there is no edge between $R_1$ and $R_2$. Since $G$, as a prime graph, is connected, there is a path between any vertex in $R_1$ and any vertex in $R_2$ but then there is a $P_5$ in $G$ which is a contradiction showing (\ref{edgeRiRj}).  
\hfill $\diamond$ 

\medskip

Now assume that $R$ is not a clique in $G$ (otherwise $G$ is $2P_2$-free and thus a thin spider by Theorem \ref{ED2P2fr}). We claim that for any three distinct indexes $i,j,\ell$ and any three vertices $x \in R_i, y \in R_j, z \in R_{\ell}$, we have: 
\begin{equation}\label{nonedgedisting}
\mbox{If } xy \notin E \mbox{ then } zx \notin E \mbox{ or } zy \notin E.
\end{equation}

{\em Proof.} Suppose that for $x \in R_1, y \in R_2, z \in R_3$, $xy \notin E$ but $zx \in E, zy \in E$. Then $d_1,x,z,y,d_2$ induce a $P_5$ which is a contradiction showing (\ref{nonedgedisting}).  
\hfill $\diamond$ 

\medskip    

This also means: If $zx \in E$ and $zy \in E$ then $xy \in E$. 

\medskip  

Finally note that: 
\begin{equation}\label{RiRjneighb}
\mbox{Every vertex } x \in R_i \mbox{ has a neighbor in some } R_j, j \neq i.
\end{equation}
   
Otherwise, $x$ and $d_i$ would be true twins (which is impossible in a prime graph). 

\medskip  

Now we claim:  
\begin{lemma}\label{2EDvP5fr} 
For a prime $P_5$-free graph $G$ which is not a thin spider, any e.d. has two vertices.
\end{lemma} 

{\bf Proof.}
Suppose that $G$ is not a thin spider and thus, there is a non-edge in $R$ as described above; let $x_1 \in R_1$ and $y_1 \in R_2$ with $x_1y_1 \notin E$. 
Let $D=\{d_1,d_2,\ldots,d_k\}$, be an e.d. of $G$, and assume that $k \ge 3$. Recall that by (\ref{RiRjneighb}), $x_1$ has a neighbor in another $R_j$, $j \neq 1$, and $y_1$ has a neighbor in another $R_{\ell}$, $\ell \neq 2$. 

\medskip

{\bf Case 1.} First let $x_1$ have a neighbor $y_2 \in R_2$, and let $y_1$ have a neighbor $x_2 \in R_1$. Then, by (\ref{edgeRiRj}), there is an edge between $R_1$ and $R_3$, and there is an edge between $R_2$ and $R_3$; let $y \in R_2$ and $z \in R_3$ with $yz \in E$. By (\ref{nonedgedisting}), we have $zx_1 \notin E$ or 
$zy_1 \notin E$. First assume that $zx_1 \notin E$. Then by (\ref{nonedgedisting}), $x_1y \notin E$ and $zy_2 \notin E$, and now $x_1,y_2,y,z,d_3$ induce a $P_5$ which is a  contradiction. Now assume that $zy_1 \notin E$. Then by (\ref{nonedgedisting}), $zx_2 \notin E$ which implies $x_2y \notin E$, and now $x_2,y_1,y,z,d_3$ induce a $P_5$ which is a contradiction. Thus, Case 1 is excluded.  

\medskip

{\bf Case 2.} Now assume that $x_1$ has a neighbor $y_2 \in R_2$ but $y_1$ has no neighbor in $R_1$; let $v \in R_3$ be a neighbor of $y_1$. Then, by (\ref{nonedgedisting}), $x_1v \notin E$. Since $d_3,v,y_1,y_2,x_1$ does not induce a $P_5$, we have $vy_2 \in E$ which contradicts (\ref{nonedgedisting}). 
 
\medskip

{\bf Case 3.} $x_1$ has no neighbor in $R_2$, and $y_1$ has no neighbor in $R_1$. Let $u \in R_i$, $i \neq 1,2$, be a neighbor of $x_1$, and let $v \in R_j$, $j \neq 1,2$, be a neighbor of $y_1$. By (\ref{nonedgedisting}), $u \neq v$. Then, by the connectedness of $G$, we obtain a $P_5$ in any case.  
\qed

\medskip

Thus, for prime $P_5$-free graphs with e.d., we have only the following two cases: 

\begin{enumerate}
\item $G$ is a thin spider (in which case $G$ is $2P_2$-free and an e.d. $D$ can have arbitrary size). 
\item $G$ is not a thin spider; any e.d. of $G$ has two vertices. 
\end{enumerate}

In particular, we have: For a prime $P_5$-free graph $G$, an e.d. of $G$ has at least three vertices only if $G$ is a thin spider. 

\begin{lemma}\label{2EDvP5frmindeg} 
If $D=\{d_1,d_2\}$ is an e.d. of the prime $P_5$-free graph $G$ which is not a thin spider then either $deg(d_1)=deg(d_2)=\delta(G)$ and $d_1,d_2$ are the only vertices with minimum degree in $G$ or $deg(d_1)=\delta(G)$ and $d_1$ is the only vertex with minimum degree in $G$.
\end{lemma} 

{\bf Proof.}
As before, let $R_i=N(d_i)$ for $i=1,2$; we know already that $R_i$ is a clique and by (\ref{RiRjneighb}), every vertex in $R_i$ has a neighbor in $R_j$ for $i \neq j$. Thus, if $|R_1|=|R_2|=k$ then $deg(d_1)=deg(d_2)=k=\delta(G)$ and for every $x \in R_1 \cup R_2$, $deg(x) > k$. If $k=|R_1| < |R_2|$ then $deg(d_1)=k=\delta(G)$ and $d_1$ is the only vertex with minimum degree $\delta(G)$ in $G$.    
\qed

\section{The algorithm}

We give a weakly robust algorithm for the ED problem on $P_5$-free graphs (robustness is meant in the sense of Spinrad \cite{Spinr2003}): 

\medskip

{\bf Algorithm ED for $P_5$-free graphs}

\medskip

{\bf Given:} A prime graph $G=(V,E)$.\\
{\bf Question:} Does $G$ have an e.d.?

\begin{itemize}
\item[(0)] For all $v \in V$, determine the degree $deg(v)$. 
\item[(1)] Check whether $G$ is a thin spider such that $D$ is the set of all degree 1 vertices; if YES then $G$ has the uniquely determined e.d. $D$ - STOP.
\item[(2)] $\{$Now $G$ is not a thin spider.$\}$ Determine a vertex $d$ of minimum degree $\delta(G)$ in $G$ and check whether $N(d)$ is a clique. If NOT then STOP - $G$ is not $P_5$-free or has no e.d. Otherwise determine a vertex $d'$ of minimum degree in $G$ among all vertices in $V \setminus N[d]$ and check whether $N(d')$ is a clique. If NOT then STOP - $G$ is not $P_5$-free or has no e.d. Otherwise, $\{d,d'\}$ is the uniquely determined e.d. of $G$. 
\item[(3)] Check if $\{d,d'\}$ is an e.d. of $G$. If yes then $G$ has e.d. $\{d,d'\}$; otherwise, $G$ has no e.d. 
\end{itemize}

\begin{theorem}\label{EDP5frlincorr} 
Algorithm ED for $P_5$-free graphs is correct and works in linear time. 
\end{theorem} 

{\bf Proof.}
{\em Correctness:} Assume that $G$ is prime and has an e.d. $D$. By the result of \cite{BraGiaMilNev2014} we know that if $G$ is prime and $2P_2$-free and has an e.d. then $G$ is a thin spider and $D$ is the set of vertices in $G$ of degree 1. Thus, the other case is that $G$ is not $2P_2$-free (but $P_5$-free). Then by 
Lemma~\ref{2EDvP5fr}, any e.d. $D$ of $G$ has two vertices, say $D=\{d,d'\}$, and by (\ref{Ndiclique}), the neighborhoods of $d$ and of $d'$ are cliques. 
By Lemma~\ref{2EDvP5frmindeg}, the only candidates for an e.d. are either the two vertices $d,d'$ with degree $\delta(G)$ or the single vertex $d$ with degree $\delta(G)$. Obviously, if $d$ is the only vertex of minimum degree in $G$ then $d'$ is the vertex of minimum degree among all vertices in $V \setminus N[d]$.    
 
\medskip

{\em Time bound:} Obviously, all steps $(0)-(3)$ can be done in linear time.  
\qed

\medskip

Since in both cases, an e.d. is uniquely determined, the algorithm also solves the WED problem. 

\medskip

{\bf Acknowledgement.}
The author thanks Martin Milani\v c for discussions and comments about this manuscript.

\begin{footnotesize}

\end{footnotesize}

\end{document}